# On the Inhibition of COVID-19 Protease by Indian Herbal Plants: An *In Silico* Investigation

Ambrish Kumar Srivastava[1, 3*], Abhishek Kumar[2, 3*], Neeraj Misra[2]

[1]*Department of Physics, Deen Dayal Upadhyaya Gorakhpur University, Gorakhpur (U.P.) India*

[2]*Department of Physics, University of Lucknow, Lucknow (U.P.) India*

[3]Authors have equal contribution

[*]Corresponding authors: ambrish.phy@ddugu.ac.in; aks.ddugu@gmail.com (A. K. Srivastava)

abhishekphy91@gmail.com (A. Kumar)

**Abstract**

COVID-19 has quickly spread across the globe, becoming a pandemic. This disease has a variable impact in different countries depending on their cultural norms, mitigation efforts and health infrastructure. In India, a majority of people rely upon traditional Indian medicine to treat human maladies due to less-cost, easier availability and without any side-effect. These medicines are made by herbal plants. This study aims to assess the Indian herbal plants in the pursuit of potential COVID-19 inhibitors using *in silico* approaches. We have considered 18 extracted compounds of 11 different species of these plants. Our calculated lipophilicity, aqueous solubility and binding affinity of the extracted compounds suggest that the inhibition potentials in the order; harsingar > aloe vera > giloy > turmeric > neem > ashwagandha > red onion > tulsi > cannabis > black pepper. On comparing the binding affinity with hydroxychloroquine, we note that the inhibition potentials of the extracts of harsingar, aloe vera and giloy are very promising. Therefore, we believe that these findings will open further possibilities and accelerate the works towards finding an antidote for this malady.

**Introduction**

The coronavirus disease (COVID-19) has been declared as a worldwide epidemic by the world health organization (WHO). According to the latest update of the WHO[1], there are more than 1 million cases of the COVID-19 worldwide, causing almost 50 thousand deaths affecting 203 countries, areas or territories. This novel coronavirus was detected in late December 2019 and recognized in early January 2020 in China[2,3]. On 11 February 2020, the international committee on taxonomy of viruses declared this "severe acute respiratory syndrome coronavirus 2" (SARS-CoV-2) as the new virus[4]. Presumably, this virus picks up its name from the virus responsible for the SARS outbreak of 2003 as they are genetically related but different. Subsequently, the WHO announced "COVID-19" as the name of this new disease[5]. The COVID-19 spread abruptly across the globe, becoming a pandemic within a couple of weeks and apart from China, the news of the carnage started pouring in from countries like the USA, Italy, Spain, France, Germany and Iran along with India on a day to day basis. After the first case of the COVID-19 on 30 January 2020 in India, there become 1466 cases and 38 deaths as per official data of the ministry of health and family welfare, India[6].

Unfortunately, there has been no noticeable breakthrough in the management of this disease to date and the patient is given a treatment based on his observable and diagnosable symptoms. Although several attempts have been made in the research and development of the diagnostics, therapeutics and vaccines for this novel coronavirus[7], there exists no chemotherapeutic agent so far which has been shown unequivocally to be effective in treating human diseases due to a minuscule virus. To combat this deadly COVID-19, a number of conventional drugs[8-10] like chloroquine, hydroxychloroquine, remdesivir, etc. have been tried and found with certain curative effect *in vitro*. However, the clinical drug response is not very encouraging and toxicity remains an inevitable issue causing serious adverse effects[11]. This prompted us to study the inhibition of COVID-19 protease by Indian herbal plants.

Because of the inherent side effects of the synthetic chemicals used in allopathic drugs, a sizeable population has switched over to the traditional system of medicine (herbal medicine) for their primary health care. Ayurveda, the age-old Indian system of medicine, is increasingly becoming a sought after system to bank on. The ayurvedic treatment has become an alternative to conventional medicines due to several reasons including easy availability, less or no side effects and less cost. India has always been a rich reservoir of medicinal plants because of several agro-climatic zones. Therefore, in the present work, we have chosen a multitude of Indian herbal plants such as harsingar (*Nyctanthes arbor-tristis*), giloy (*Tinospora cordifolia*), aloe vera (*Aloe barbadensis miller*), turmeric (*Curcuma longa*), neem (*Azadirachta indica*), ashwagandha (*Withania somnifera*), ginger (*Zingiber officinale*), red onion (*Allium cepa*), tulsi (*Ocimum sanctum*), cannabis (*Cannabis sativa*) and black pepper (*Piper nigrum*). The pharmacological importance of these plants is well documented in the literature[12-14]. We have selected a few extracted compounds of these herbal plants and evaluated their inhibition properties against COVID-19 main protease *in silico*. We have obtained encouraging responses from most of these medicinal plants in general. The inhibition potentials of harsingar, aloe vera and giloy are particularly interesting. Therefore, we believe that this study should offer some insights into the development of alternative drugs for this novel coronavirus.

**Methodology**

This study was performed by the SwissDock web server[15,16], which incorporates an automated *in silico* molecular docking procedure based on the EADock ESS docking algorithm. To determine the inhibition properties of Indian medicinal plants against COVID-19 protease, the potential target protein was retrieved from the RCSB protein data bank (PDB ID: 6LU7) deposited in February 2020[17]. The processed coordinates files for the ligands as well as COVID-19 protease (6LU7) has been uploaded, and docking was carried out with the 'Accurate' parameter option, which is considered to be the most extensive for the sampling of the binding modes. The output clusters have been obtained after each docking runs and classified based on the full fitness (FF) score by the SwissDock algorithm. A greater negative FF score suggests a more favorable binding mode between ligand and receptor with a better fit. The visual graphics of docking results have been generated by using the UCSF Chimera program[18].

In addition, we have calculated the lipophilicity (log $P$) and aqueous solubility (log $S$) using ALOGPS 2.1 program[19], which is based on the electro-topological state indices and associative neural network modeling[20]. These two parameters are very important for quantitative structure-property relationship (QSPR) studies.

**Results and Discussion**

We have considered a total of 11 different varieties (species) of Indian medicinal plants. Fig. 1 displays the molecular structures of a few (main) compounds extracted from these plants. We have focused on mainly those compounds which have been found to possess anti-malarial, anti-viral or other similar activities. Nictoflorin ($C_{27}H_{30}O_{15}$), astragalin ($C_{21}H_{20}O_{11}$), lupeol ($C_{25}H_{26}O_4$) are extracted from the leaves of harsingar. Berberine ($C_{28}H_{18}NO_4$) and sitosterol ($C_{29}H_{50}O$) are chemical constituents of the stem of the giloy. Aloenin ($C_{19}H_{22}O_{10}$) and aloesin ($C_{19}H_{22}O_9$) are extracted from aloe vera leaves. Curcumin ($C_{21}H_{20}O_6$) is extracted from the dried ground rhizome of the turmeric. Nimbin ($C_{30}H_{36}O_9$) is the first bitter compound isolated from the oil of neem. Withanolide ($C_{28}H_{38}O_6$) and withaferin A ($C_{28}H_{38}O_6$) are steroidal constituents of ashwagandha.

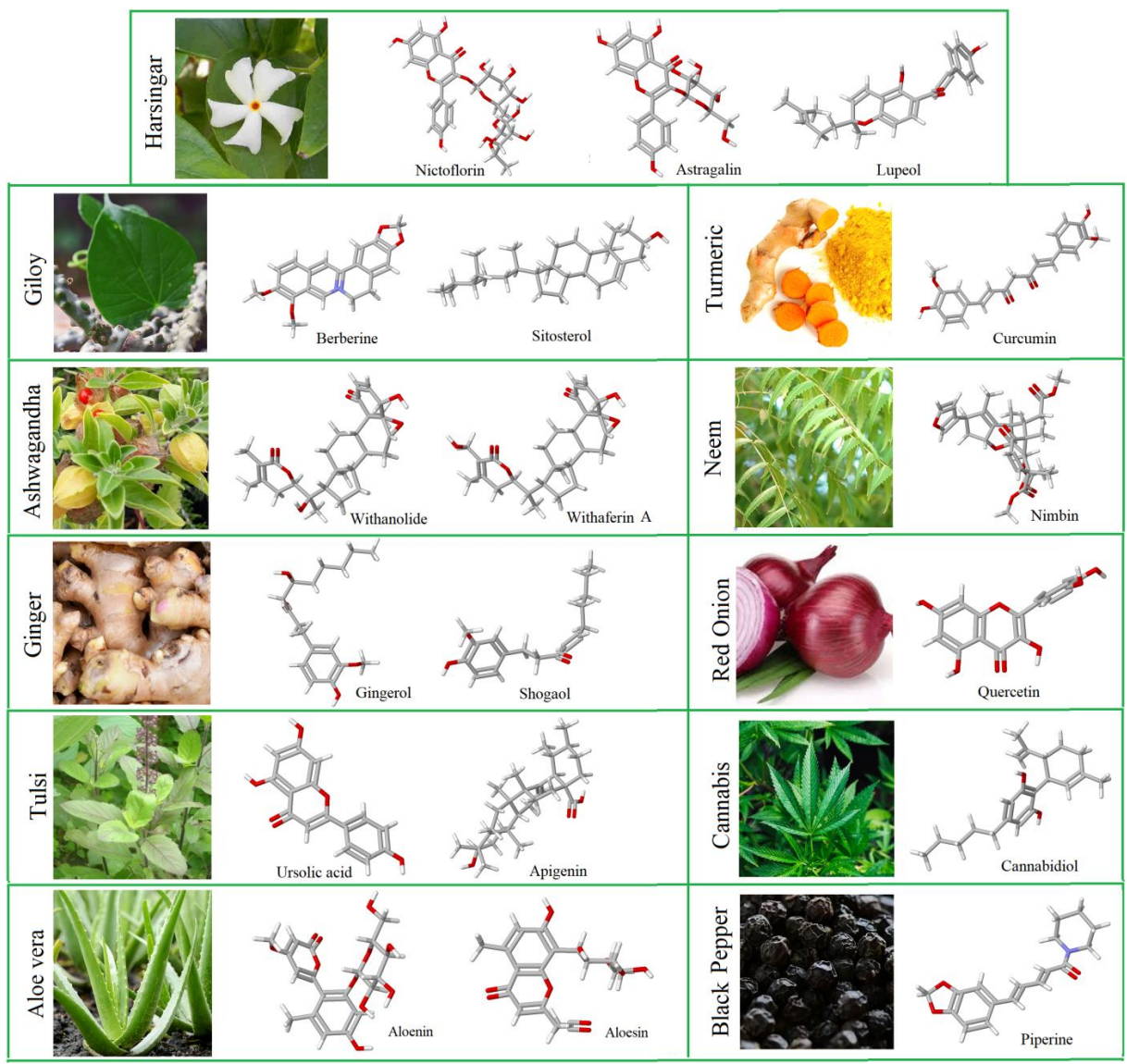

Fig. 1. Molecular structures of the compounds extracted from Indian herbal plants. Harsingar (night jasmine or parijat) is distributed widely in sub-Himalayan regions, southwards to Godavari and also found in Indian gardens as ornamental plant. Giloy (moonseed plant or guduchi) is a large deciduous, extensively spreading climbing shrub found throughout India and also in Bangladesh, Srilanka and China. Aloe vera (ghrit kumari) is a well known medicinal plant with sharp pointed, lanced shaped and edged leaves having its origin in African content, Turmeric (circumin or haldi), a traditional Chinese medicine, is commonly used species in Indian subcontinent, not only for health but also for the preservation of food. Ashwagandha is known as Indian winter cherry. Neem (margosa tree) also called as Indian lilac with its centre of origin in southern and southeastern Asia, is regarded as 'village dispensary' in India and also a religious gift from nature. Red onion is a versatile vegetable, i.e., consumed fresh as well as in the form of processed products. Tulsi is the one of the most religious and medicinal plant in India and grown throughout the country from Andaman and Nicobar island to the Himalayas. Cannabis is a plant of psychoactive drug and black pepper is a kind of household species used in India.

Gingerol ($C_{17}H_{26}O_4$) and shogaol ($C_{17}H_{24}O_3$) are two constituents of pungent ketones, which result in the strong aroma of ginger. Quercetin ($C_{15}H_{10}O_7$) is the main flavonoid content of (red) onion. Ursolic acid ($C_{30}H_{48}O_3$) and apigenin ($C_{15}H_{10}O_5$) are chemical constituents of tulsi leaves. Cannabidiol ($C_{21}H_{30}O_2$) major constituent of cannabis extracts and is devoid of the typical psychological effects of cannabis in humans. Piperine ($C_{17}H_{19}NO_3$) is a naturally occurring alkaloid, was isolated from the plants of both the black and white pepper grains.

In order to compare the biological activity and pharmacological behavior of the extracted compounds, we have evaluated their log $P$ as well as log $S$ values and listed in Table 1. Log $P$ measures the hydrophilicity of a compound. The compounds having high log $P$ values show poor absorption or low permeability. One can note that the log $P$ values of most of these compounds lie in the range 2.64-4.95. These values indicate that the compounds can easily diffuse across the cell membranes due to their high organic (lipid) permeability. However, lupeol, sitosterol, ursolic acid and cannabidiol have log $P$ in the range 5.12-7.27 (exceeding to 5) and therefore, they possess high hydrophobicity and poor absorption. On the contrary, nictoflorin, astragalin, aloenin, aloesin and quercetin possess high absorption due to their log $P$ in the range 0.05-1.81. Log $S$ represents the aqueous solubility of the compound. It is an important factor, associated with the bioavailability of compounds. Most of these compounds have log $S$ values higher than -5, except lupeol, sitosterol and ursolic acid. Note that the log $S$ values of more than 85% of compounds (drugs) fall in the range between -1 and -5 [21]. This is consistent with their log $P$ values as poor solubility implies poor absorption and hence, bioavailability. Thus, log $P$ along with log $S$ values of these compounds confirmed their permeability across cell membranes. In particular, the nictoflorin, astragalin, aloenin, aloesin and quercetin seem to be more biologically potent. These parameters are also associated with their interaction with receptors.

Table 1. Calculated parameters of compounds extracted from Indian herbal plants as possible inhibitors of COVID-19 protease.

| Indian herbal plants | Extracted compounds | Log $P$ | Log $S$ | Binding affinity (kcal/mol) | FF score | Active sites/Binding residue/ H-bond length (Å) |
|---|---|---|---|---|---|---|
| Harsingar | Nictoflorin ($C_{27}H_{30}O_{15}$) | 0.07 | -2.29 | -9.18 | -1057 | N-H--O/GLY-143/2.311 |
|  | Astragalin ($C_{21}H_{20}O_{11}$) | 0.52 | -2.45 | -8.68 | -1123 | O--H/PHE-140/2.197 |
|  | Lupeol ($C_{25}H_{26}O_4$) | 5.12 | -5.26 | -8.28 | -1160 | N-H--O/THR-26/2.027 |
| Aloe vera | Aloenin ($C_{19}H_{22}O_{10}$) | 0.05 | -2.35 | -9.13 | -1120 | O--H/PHE-140/2.151 |
|  | Aloesin ($C_{19}H_{22}O_9$) | 0.12 | -1.99 | -8.79 | -1135 | N-H--O/GLY-143/2.016 N-H--O/GLU-166/2.297 |
| Giloy | Berberine ($C_{28}H_{18}NO_4$) | 3.75 | -4.16 | -8.67 | -1168 | N-H--O/GLY-143/2.540 N-H--O/GLY-143/2.577 |
|  | Sitosterol ($C_{29}H_{50}O$) | 7.27 | -7.35 | -8.42 | -1178 | O--H/PHE-166/2.080 |
| Turmeric | Curcumin ($C_{21}H_{20}O_6$) | 3.62 | -4.81 | -8.44 | -1196 | N-H--O/GLY-143/2.243 |
| Neem | Nimbin ($C_{30}H_{36}O_9$) | 3.71 | -4.36 | -8.17 | -1128 | N-H--O/GLY-143/2.161 |
| Ashwagandha | Withanolide ($C_{28}H_{38}O_6$) | 2.70 | -4.91 | -8.07 | -915 | O--H/GLU-166/1.991 N-H--O/GLU-166/2.110 |
|  | Withaferin A ($C_{28}H_{38}O_6$) | 2.64 | -4.81 | -8.05 | -942 | N-H--O/GLY-143/2.577 |
| Ginger | Gingerol ($C_{17}H_{26}O_4$) | 3.45 | -3.57 | -7.95 | -1220 | O--H/THR-190/2.026 |
|  | Shogaol ($C_{17}H_{24}O_3$) | 4.95 | -4.49 | -7.86 | -1209 | N-H--O/GLY-143/2.289 N-H--O/THR-26/2.398 O--H/THR-24/2.345 |
| Red Onion | Quercetin ($C_{15}H_{10}O_7$) | 1.81 | -3.06 | -7.70 | -1189 | O--H/THR-26/1.936 |
| Tulsi | Ursolic acid ($C_{30}H_{48}O_3$) | 6.35 | -5.89 | -7.46 | -1152 | N-H--O/GLY-143/2.330 |
|  | Apigenin ($C_{15}H_{10}O_5$) | 3.07 | -3.36 | -7.38 | -1210 | O--H/THR-26/1.994 |
| Cannabis | Cannabidiol ($C_{21}H_{30}O_2$) | 6.10 | -4.40 | -7.10 | -1214 | N-H--O/GLY-143/2.325 |
| Black Pepper | Piperine ($C_{17}H_{19}NO_3$) | 3.38 | -3.28 | -6.98 | -1211 | N-H--O/THR-26/2.529 |

The molecular docking studies explore the interaction mechanism between ligands and receptors. The interactions between a ligand and receptor play a crucial role in the field of drug discovery. The molecular docking calculations have been performed as blind, i.e., covered the entire protein surface, not any specific region of the protein as the binding pocket in order to avoid sampling bias. The docking parameters such as binding affinity, FF score, and H-bond, bond-length along with amino acids (residue) found in the binding site pockets (active site) of 6LU7 are listed in Table 1. The results of molecular docking are displayed in Fig. 2.

The binding affinity ($\Delta G$) of (drug) compounds depends on the type of bonding (H-bond) that occurs with the active site of the protein. The results of docking show that the extracts of harsingar, nictoflorin, astragalin and lupeol form H-bond of bond lengths 2.311 Å, 2.197 Å and 2.027 Å with the glycine (GLY-143), phenylalanine (PHE-140) and threonine (THR-26) respectively. The compounds of aloe vera, aloenin forms H-bond (bond length = 2.151 Å) with phenylalanine (PHE-140) and aloesin forms two H-bonds with GLY-143 (bond length = 2.016 Å) and glutamate (GLU-166) with bond length = 2.297 Å. The constituents of giloy, berberine forms two H-bonds with the same amino acid GLY-143 having bond lengths of 2.540 Å and 2.577 Å, whereas sitosterol forms H-bond with PHE-166 of bond length 2.080 Å. The compounds of turmeric and neem, curcumin and nimbin form H-bonds with the same amino acid GLY-143 of bond lengths 2.243 Å and 2.161 Å, respectively. The derivatives of ashwagandha, withanolide forms two H-bonds with the GLU-166 (bond length = 1.991 Å and 2.110 Å) and withaferin A forms H-bond with GLY-143 of bond length 2.577 Å. The compounds of ginger, gingerol forms H-bond with THR-190 (bond length = 2.026 Å) and shogaol forms three H-bonds, one with GLY-143 (bond length = 2.016 Å) and two with the THR-26 and THR-24 having bond lengths 2.398 Å and 2.345 Å, respectively.

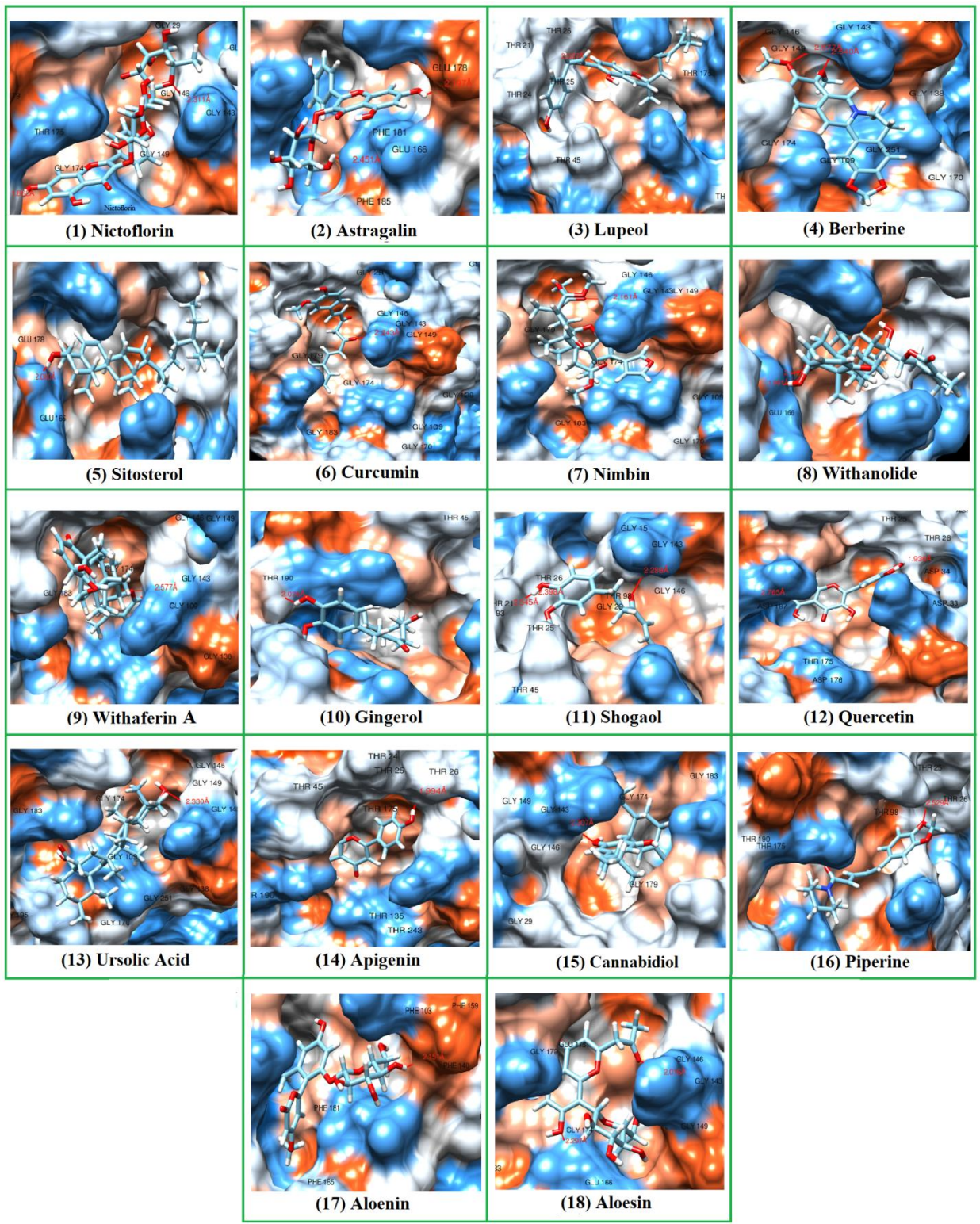

Fig. 2. The binding of COVID-19 protease (6LU7) receptor with the extracts of Indian herbal plants explored by molecular docking.

The derivatives of red onion and black pepper, quercetin and piperine form H-bond with the THR-26 of bond lengths 2.243 Å and 2.161 Å, respectively. The extracts of tulsi, ursolic acid and apigenin form H-bonds with the GLY-143 (bond length = 2.330 Å) and the THR-26 (bond length = 1.994 Å), respectively. The constituent of cannabis, cannabidiol forms H-bond with the GLY-143 (bond length = 2.325 Å), a non-polar amino acid.

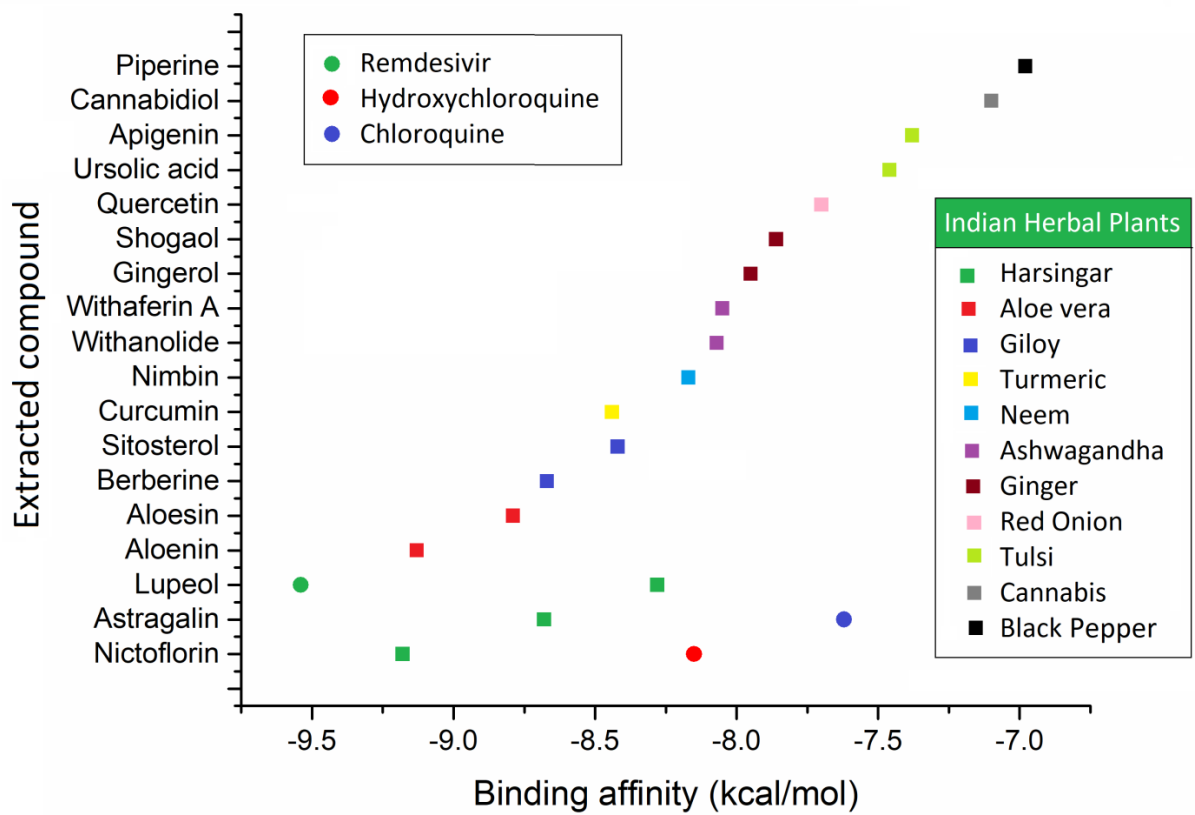

Fig. 3. Binding affinity plot of Indian herbal plants and a few drugs for comparison of inhibition potential against COVID-19 protease.

Thus, our docking analyses suggest that the COVID-19 protease (6LU7) can be inhibited by the extracts of Indian herbal plants. Based on the binding affinity, the inhibition potential of these plants (based on their extracts) can be ranked as; harsingar > aloe vera > giloy > turmeric > neem > ashwagandha > red onion > tulsi > cannabis > black pepper. The highest inhibition potentials are obtained for the extracts of harsingar and aloe vera, namely nictoflorin (ΔG = - 9.18) and aloenin ($\Delta G$ = -9.13), respectively. This also provides us an opportunity to compare the $\Delta G$ value of the compounds extracted from other plants. Fig. 3 plots the binding affinities of these compounds, along with those of a few previously reported inhibitors such as remdesivir, chloroquine and hydroxychloroquine. Considering hydroxychloroquine as a reference, we note that the inhibition potentials of the extracts of harsingar, aloe vera and giloy are very encouraging. The extracts of turmeric, neem, ashwagandha and ginger have larger inhibition potentials than that of chloroquine. The compounds extracted from other plants also possess certain inhibition properties against COVID-19 protease.

**Conclusions and Perspectives**

We have performed an *in silico* study on the inhibition of COVID-19 protease by the extracts of Indian herbal plants. We noticed that all these plants possess inhibition properties to a certain extent. Based on the binding affinity as well as log *P* and log *S* values, harsingar, aloe vera and giloy appear as the most powerful inhibitors among the eleven plants considered here. Other potential inhibitors of COVID-19 protease include turmeric, neem, ashwagandha and ginger. The inhibition potentials of all these plant extracts are found to be larger than those of chloroquine and hydroxychloroquine. These two anti-malarial drug compounds are already reported to inhibit COVID-19 protease *in vitro*. Due to inherent toxicity and side-effects, however, they are not approved by most of the countries. Therefore, our findings become very interesting towards the development of alternative (herbal) medicines having no apparent side-effects. We expect prompt actions in this direction to combat with the COVID-19.